# Fast Rotation of the $N = Z$ Nucleus $^{36}$Ar

C.E. Svensson,[1] A.O. Macchiavelli,[2] A. Juodagalvis,[3] A. Poves,[4] I. Ragnarsson,[3] S. Åberg,[3] D.E. Appelbe,[5] R.A.E. Austin,[5] C. Baktash,[6] G.C. Ball,[7] M.P. Carpenter,[8] E. Caurier,[9] R.M. Clark,[2] M. Cromaz,[2] M.A. Deleplanque,[2] R.M. Diamond,[2] P. Fallon,[2] M. Furlotti,[10] A. Galindo-Uribarri,[6] R.V.F. Janssens,[8] G.J. Lane,[2] I.Y. Lee,[2] M. Lipoglavsek,[6] F. Nowacki,[11] S.D. Paul,[6] D.C. Radford,[6] D.G. Sarantites,[10] D. Seweryniak,[8] F.S. Stephens,[2] V. Tomov,[10] K. Vetter,[2] D. Ward,[2] and C.H. Yu[6]

[1]Department of Physics, University of Guelph, Guelph, Canada
[2]Nuclear Science Division, Lawrence Berkeley Laboratory, Berkeley, U.S.A.
[3]Department of Mathematical Physics, Lund University, Lund, Sweden
[4]Departamento de Física Teórica, Universidad de Madrid, Madrid, Spain
[5]Department of Physics & Astronomy, McMaster University, Hamilton, Canada
[6]Physics Division, Oak Ridge National Laboratory, Oak Ridge, U.S.A.
[7]TRIUMF, 4004 Wesbrook Mall, Vancouver, Canada
[8]Argonne National Laboratory, Argonne, U.S.A.
[9]Institut de Recherches Subatomiques, Université Louis Pasteur, Strasbourg, France
[10]Chemistry Department, Washington University, St. Louis, U.S.A.
[11]Laboratoire de Physique Theorique, Université Louis Pasteur, Strasbourg, France

A highly-deformed rotational band has been identified in the $N = Z$ nucleus $^{36}$Ar. At high spin the band is observed to its presumed termination at $I^\pi = 16^+$, while at low spin it has been firmly linked to previously known states in $^{36}$Ar. Spins, parities, and absolute excitation energies have thus been determined throughout the band. Lifetime measurements establish a large low-spin quadrupole deformation ($\beta_2 = 0.46 \pm 0.03$) and indicate a decreasing collectivity as the band termination is approached. With effectively complete spectroscopic information and a valence space large enough for significant collectivity to develop, yet small enough to be meaningfully approached from the shell model perspective, this rotational band in $^{36}$Ar provides many exciting opportunities to test and compare complementary models of collective motion in nuclei.

PACS numbers: 21.10.Re, 21.10.Tg, 21.60.Cs, 23.20.Lv, 27.30.+t

## 1. Introduction

The microscopic description of collective phenomena is a fundamental goal of quantum many-body physics. In the field of nuclear structure, a classic example is the desire to connect the deformed intrinsic states and collective rotational motion of nuclei with microscopic many-particle wavefunctions in the laboratory system. For nuclei in the lower *sd* shell, this connection is well understood in terms of Elliot's SU(3) model [1]. For heavier nuclei, however, the spin-orbit interaction removes the degeneracies of the harmonic oscillator, destroying the SU(3) symmetry. Furthermore, rotational motion in these nuclei generally involves valence spaces comprised of two major shells for both protons and neutrons [2]. Exact shell model diagonalizations in such spaces are well beyond current computational capabilities, and much effort has therefore been devoted to determining the approximate symmetries that permit a description of the collective motion in these nuclei within an appropriately truncated model space (cf. [3, 4, 5].) It is clearly desirable to test the validity of such approximate symmetries in cases where the valence space is large enough for collective motion to develop, but small enough to permit exact diagonalizations. Here we discuss the identification [6] of a superdeformed (SD) band in the $N = Z$ nucleus $^{36}$Ar which, like rotational bands in heavier nuclei, involves two major shells for both protons and neutrons, yet satisfies the above criteria.

## 2. Experiments

High-spin states in the $N = Z$ nucleus $^{36}$Ar were populated via the $^{24}$Mg($^{20}$Ne,2$\alpha$)$^{36}$Ar fusion-evaporation reaction in two experiments with an 80-MeV, $\sim$2 pnA $^{20}$Ne beam provided by the ATLAS facility at Argonne National Laboratory. In the first experiment, a self-supporting 440 $\mu$g/cm$^2$ $^{24}$Mg foil was used, while in the second experiment a 420 $\mu$g/cm$^2$ $^{24}$Mg layer was deposited onto a 11.75 mg/cm$^2$ Au backing in order to measure lifetimes by Doppler-shift attenuation methods. In both experiments, $\gamma$ rays were detected with 101 HPGe detectors of the Gammasphere array [7], in coincidence with charged particles detected and identified in the Microball [9], a 4$\pi$ array of 95 CsI(Tl) scintillators. The Hevimet collimators were removed from the BGO Compton-suppression shields of the HPGe detectors in order to provide $\gamma$-ray sum-energy and multiplicity information for each event [8]. The 2$\alpha$ evaporation channel leading to $^{36}$Ar was cleanly selected in the offline analysis by requiring the detection of 2 alpha particles in the Microball in combination with a total detected $\gamma$-ray plus charged-particle energy consistent with the $Q$-value for this channel [10]. Totals of 775 million and 827 million particle–$\gamma$–$\gamma$–$\gamma$ and higher fold coincidence events were recorded in the first and second experiments, respectively.

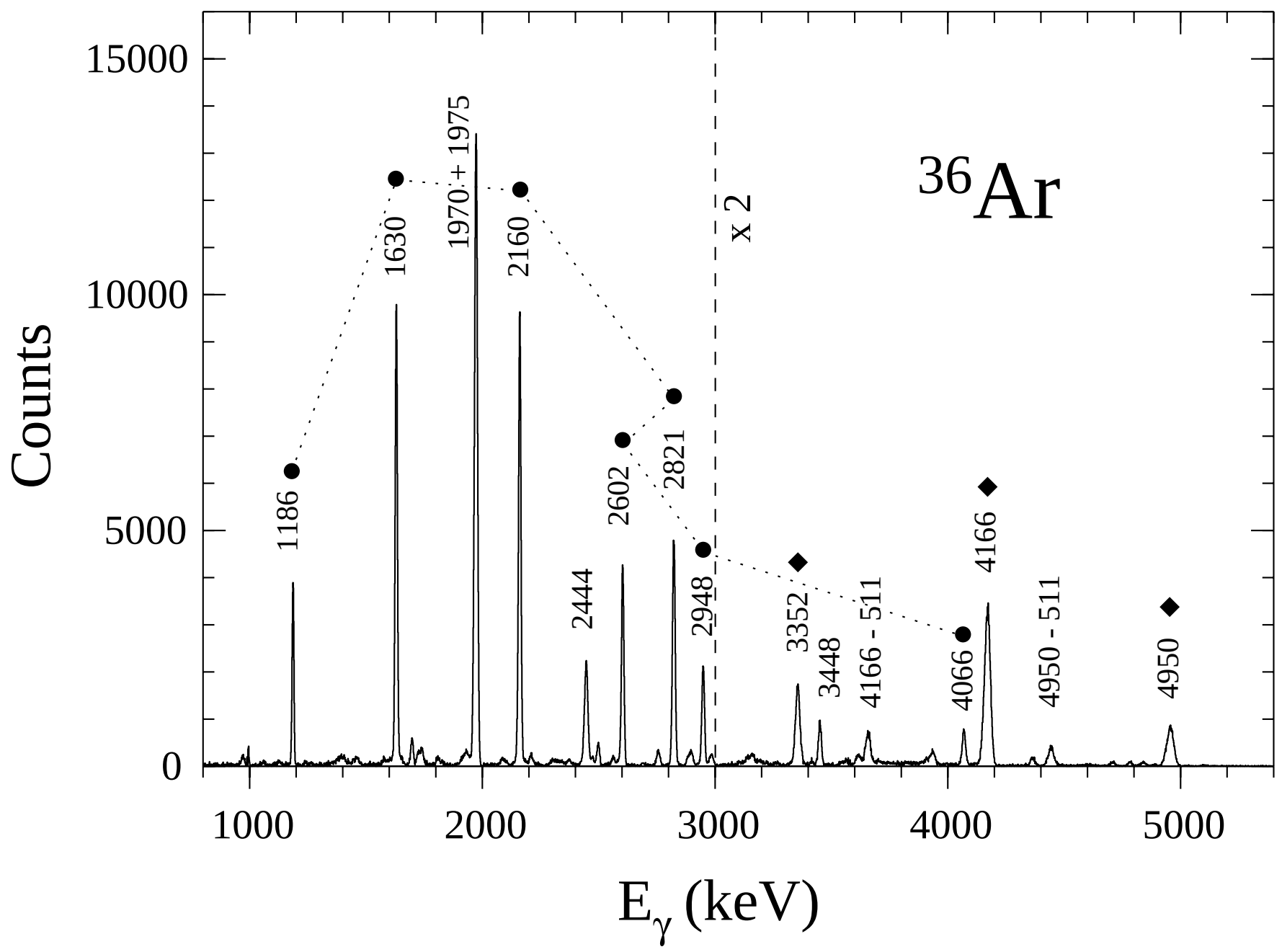

Fig. 1. Gamma-ray spectrum obtained by summing coincidence gates set on all members of the superdeformed band (circles). Diamonds indicate linking transitions connecting the band to low-spin states in $^{36}$Ar.

## 3. Results and Discussion

The $\gamma$-ray spectrum from the thin-target experiment obtained by summing coincidence gates set on all members of the superdeformed band identified in $^{36}$Ar is shown in Fig. 1, and a partial decay scheme for $^{36}$Ar is presented in Fig. 2. As shown in Fig. 2, the high-energy $\gamma$ rays indicated by diamonds in Fig. 1 link the SD band to previously known low-spin states, and thereby fix the absolute excitation energies of the SD levels. The spectrum of the SD band obtained with a single gate on the 4950 keV $\gamma$ ray, shown in the upper inset of Fig. 2, illustrates the quality of the coincidence data for these linking transitions. Angular distribution measurements establish stretched quadrupole character for the high-energy linking $\gamma$ rays and for all of the in-band transitions. The lower insets in Fig. 2 present two examples of such measurements, which determine the spins, and, under the reasonable assumption that the transitions are $E2$ rather than $M2$, the parities, of the SD states from $2^+$ to $16^+$. Although $\gamma$ decay by a 622 keV transition to the previously known $(0^+)$ state at 4329 keV [11] was not observed in this experiment, this state is presumed to be the SD bandhead based on the regular rotational spacing.

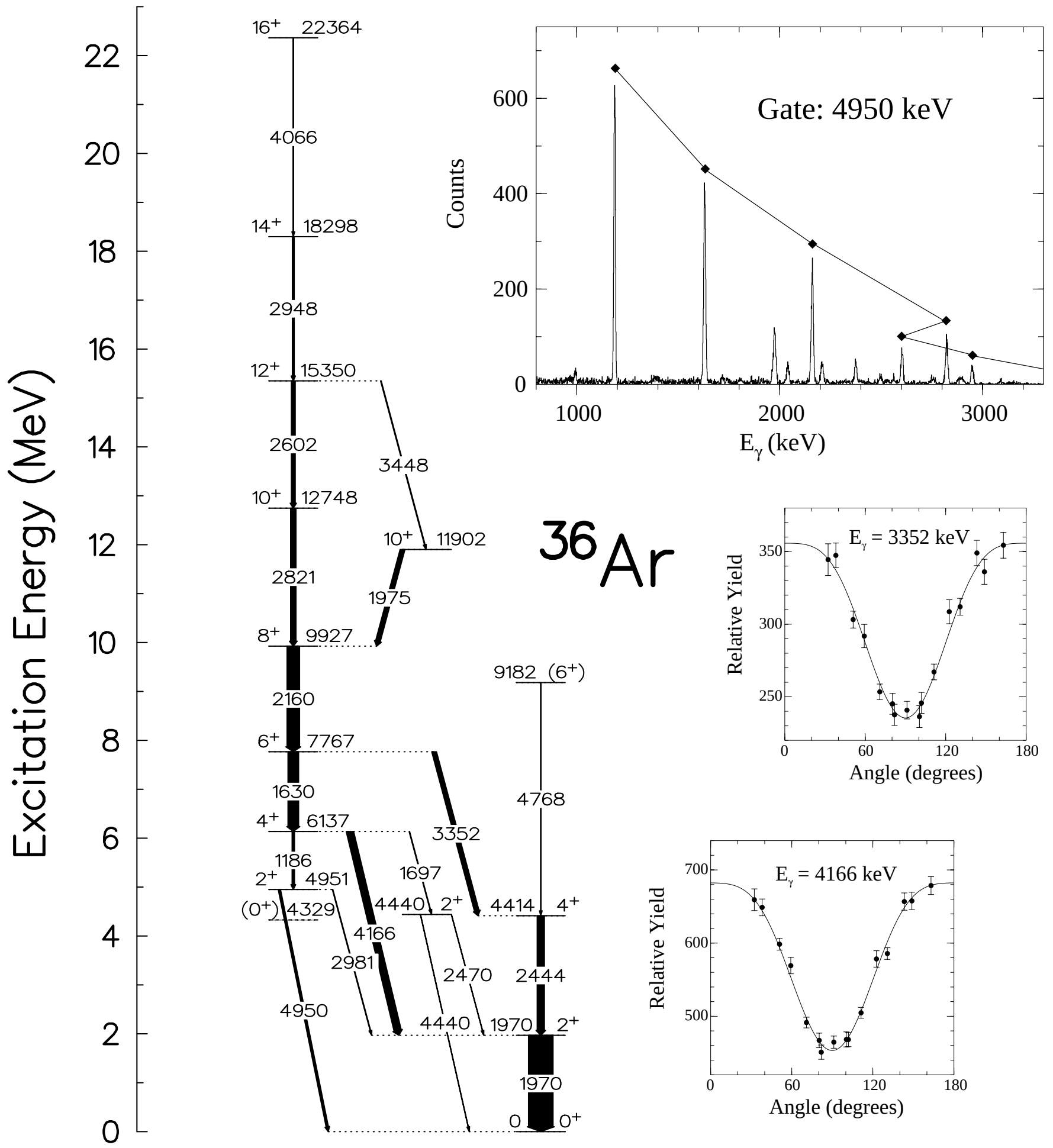

Fig. 2. Partial decay scheme for $^{36}$Ar showing the superdeformed band (left). The upper inset shows a spectrum of the band obtained with a single gate set on the 4950 keV linking transition. Angular distributions relative to the beam axis are presented for the 3352 keV and 4166 keV linking transitions in the lower insets.

We have performed both configuration-dependent [12] cranked Nilsson-Strutinsky (CNS) and large-scale $s_{1/2}d_{3/2}$-$pf$ spherical shell model (SM) calculations with the diagonalization code ANTOINE [13] for the $^{36}$Ar SD band. As detailed in Ref. [6], these calculations provide a good description of the energetic properties of the band and enable its firm assignment to a configuration in which four $pf$-shell orbitals are occupied. Here we focus primarily on the quadrupole properties of the band.

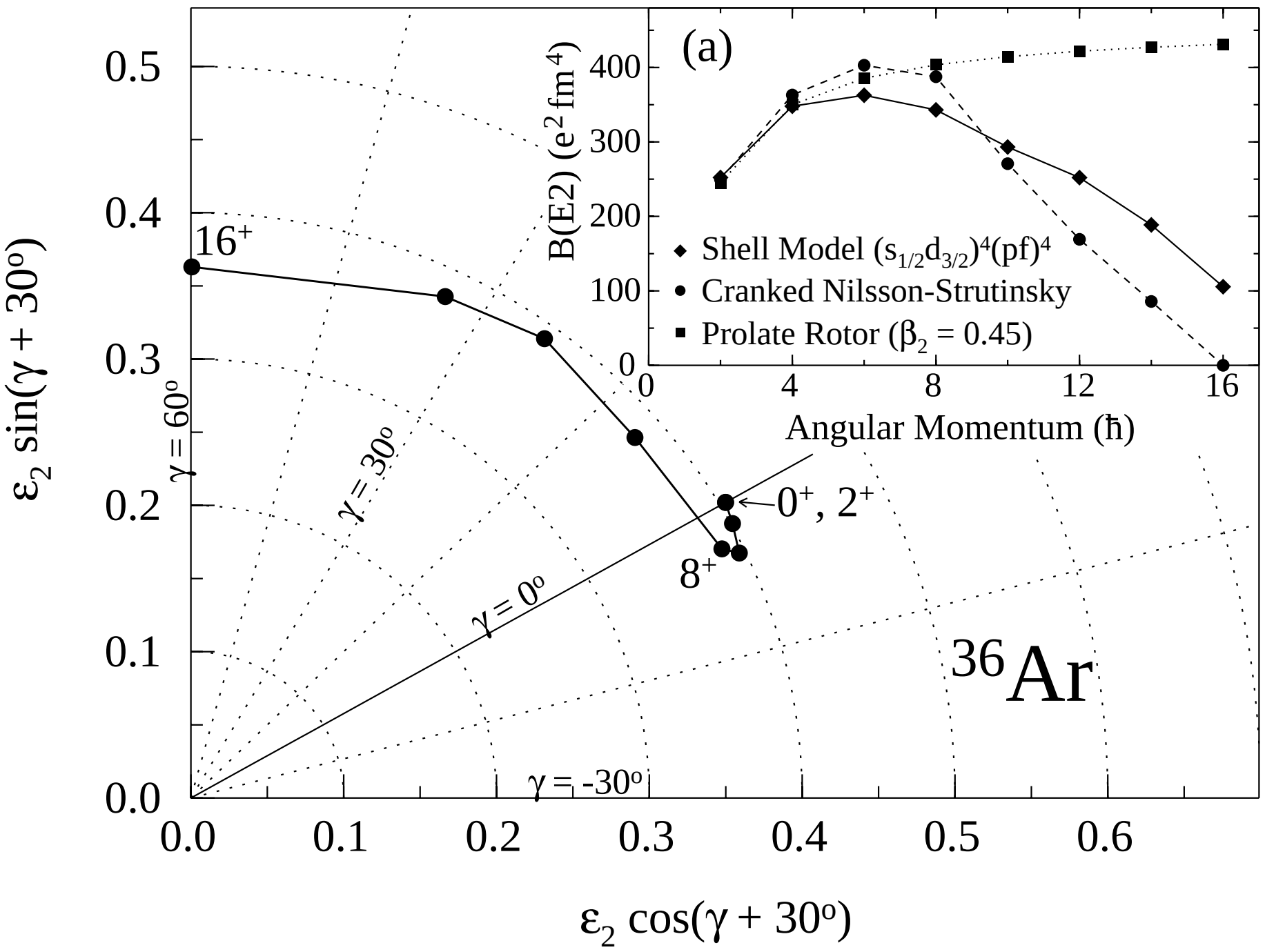

Fig. 3. Calculated trajectory in the ($\varepsilon_2$,$\gamma$) deformation plane for the SD band in $^{36}$Ar. The inset shows the corresponding $B(E2)$ values from the cranked Nilsson-Strutinsky (circles) and shell model (diamonds) calculations. The $B(E2)$'s for a fixed prolate deformation $\beta_2 = 0.45$ are also shown for comparison (squares).

The predicted equilibrium deformations for the SD band from the CNS calculations are shown in Fig. 3. At low spin, a prolate shape with deformation $\varepsilon_2 = 0.40$, which remains approximately constant up to $I = 8\hbar$, is calculated. At higher spins, the nucleus is predicted to change shape smoothly, becoming increasingly triaxial until the band terminates in a fully aligned oblate ($\gamma = 60°$) state at $I^\pi = 16^+$. Assuming the rotor model with the initial state deformations, the $B(E2; I \rightarrow I-2)$ values corresponding to this shape trajectory are shown by the circles in the inset of Fig. 3. Also shown in this Figure are the $B(E2)$ values from the SM calculations. Both calculation indicate a large collectivity at low spin which decreases as the band termination is approached. However, the shell model calculations are seen to predict somewhat smaller $B(E2)$ values for the intermediate spin states, but somewhat larger values close to termination.

In order to test the quadrupole properties shown in Fig. 3, we have measured state lifetimes throughout the $^{36}$Ar SD band by Doppler-shift attenuation techniques. These measurements rely on changes in the velocities of the recoiling $^{36}$Ar nuclei as they pass through the target and, in the case of the second experiment, the Au target backing. It is therefore essential to have a good understanding of the initial recoil velocity distribution for the events in the particle-gated data set. In the $^{24}$Mg($^{20}$Ne,$2\alpha$)$^{36}$Ar reaction, the evaporation of two alpha particles from the compound $^{44}$Ti nucleus produces a very broad recoil distribution for the $^{36}$Ar nuclei. Furthermore, the alpha-particle detection efficiency is strongly angle dependent. The alphas emitted forward (relative to the beam direction) in the center-of-mass (CM) frame receive an energy boost in the lab frame from the CM motion, and are detected more efficiently than those emitted at backward angles, which have lower energies in the lab system. A result of requiring $2\alpha$ detection is thus to bias the experimental data set towards events in which the alpha particles were emitted in the forward direction, and consequently the $^{36}$Ar recoil velocity was reduced compared to that of the compound system. This effect is clearly seen in Fig. 4, which shows the experimental recoil velocity distribution for the $^{36}$Ar nuclei reconstructed from the measured energies and directions of the detected alpha particles. In order to obtain meaningful lifetime measurements, it was necessary to modify the Lineshape code [14] to incorporate this experimentally determined initial recoil velocity distribution. The excellent fits to the $\gamma$-ray lineshapes that were obtained by this method are illustrated with the 2160 keV $8^+ \rightarrow 6^+$ SD transition in Fig. 5.

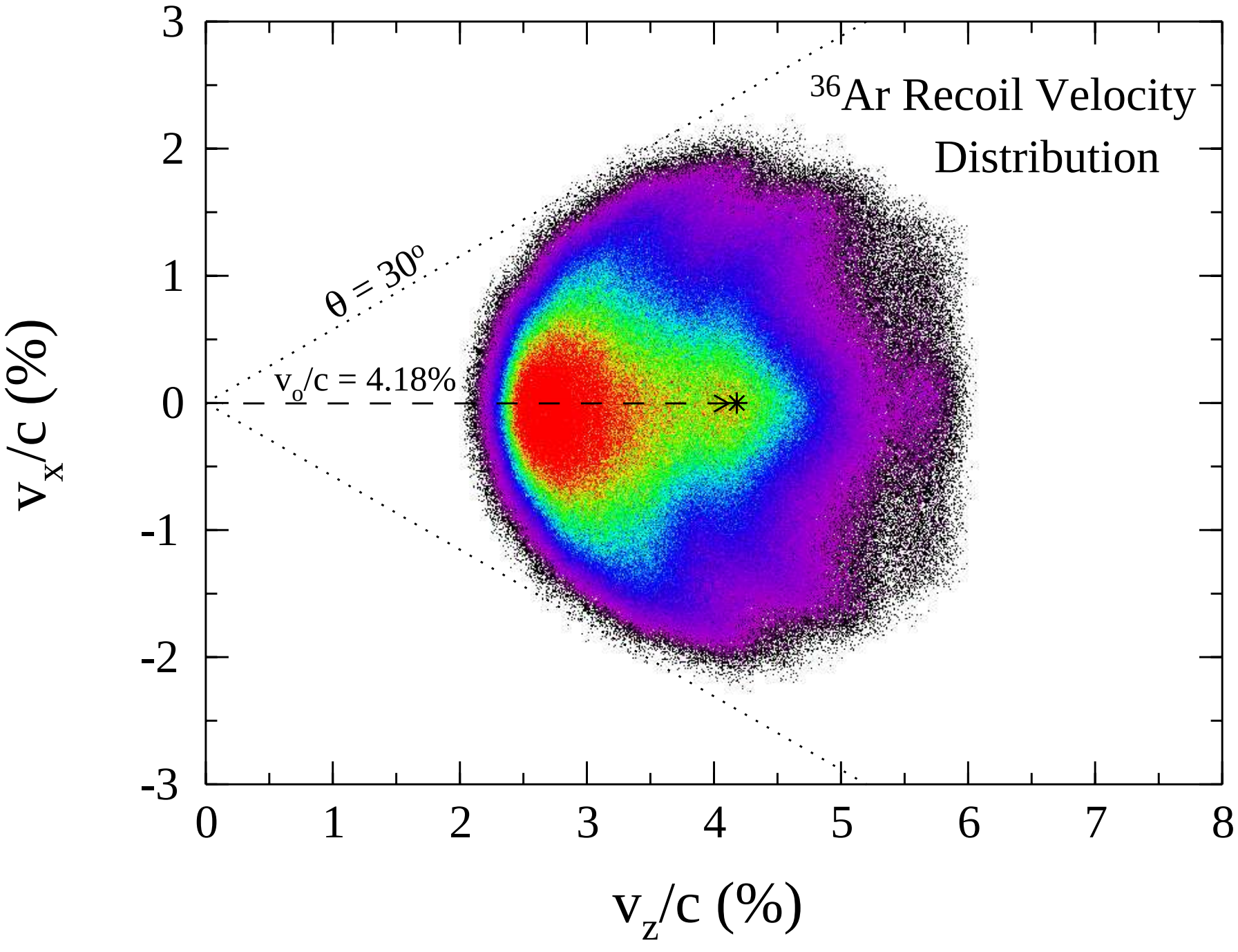

Fig. 4. Initial recoil velocity distribution in the $v_x$–$v_z$ plane for $^{36}$Ar nuclei in the $2\alpha$-gated data set, as reconstructed from the measured alpha-particle momenta. The beam is in the $z$-direction, and the recoil velocity for a compound nucleus formed at the center of the target was $v_z/c = 4.18\%$.

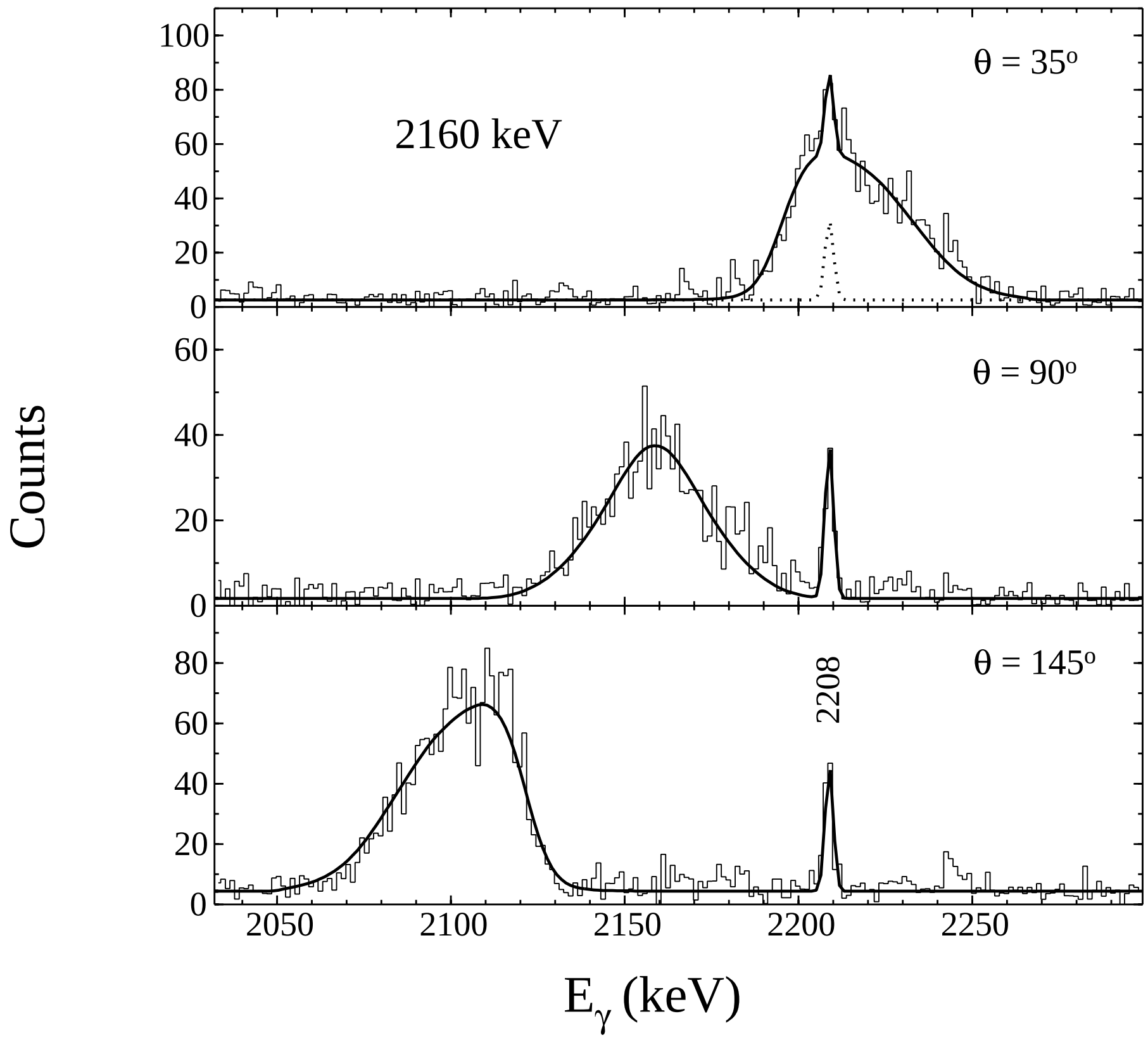

Fig. 5. Doppler-shift attenuation lineshape analysis for the 2160 keV $8^+_{SD} \rightarrow 6^+_{SD}$ transition. The stopped peak at 2208 keV is the $3^-_1 \rightarrow 2^+_1$ transition in $^{36}$Ar.

A complete description of the lifetime measurements for the $^{36}$Ar SD band is given in Ref. [15]. Here we summarize the resulting $B(E2)$ values in Fig. 6. The CNS calculations are seen to provide a good description of the measured low-spin $B(E2)$ values, but underpredict the high-spin collectivity. This result is expected as these calculations do not include the effects of shape fluctuations about the equilibrium deformations, which will remove the artificial vanishing of the $B(E2)$ at the terminating state arising from the prediction of an exactly oblate equilibrium deformation. The SM calculations provide an excellent description of the spin-dependence of the $B(E2)$ values, including the substantial remaining collectivity at the terminating state, but systematically underestimate the experimental values by approximately 20%. In order to further investigate these effects, it is interesting to compare the microscopic structures of the wavefunctions for the SD band from the two calculations.

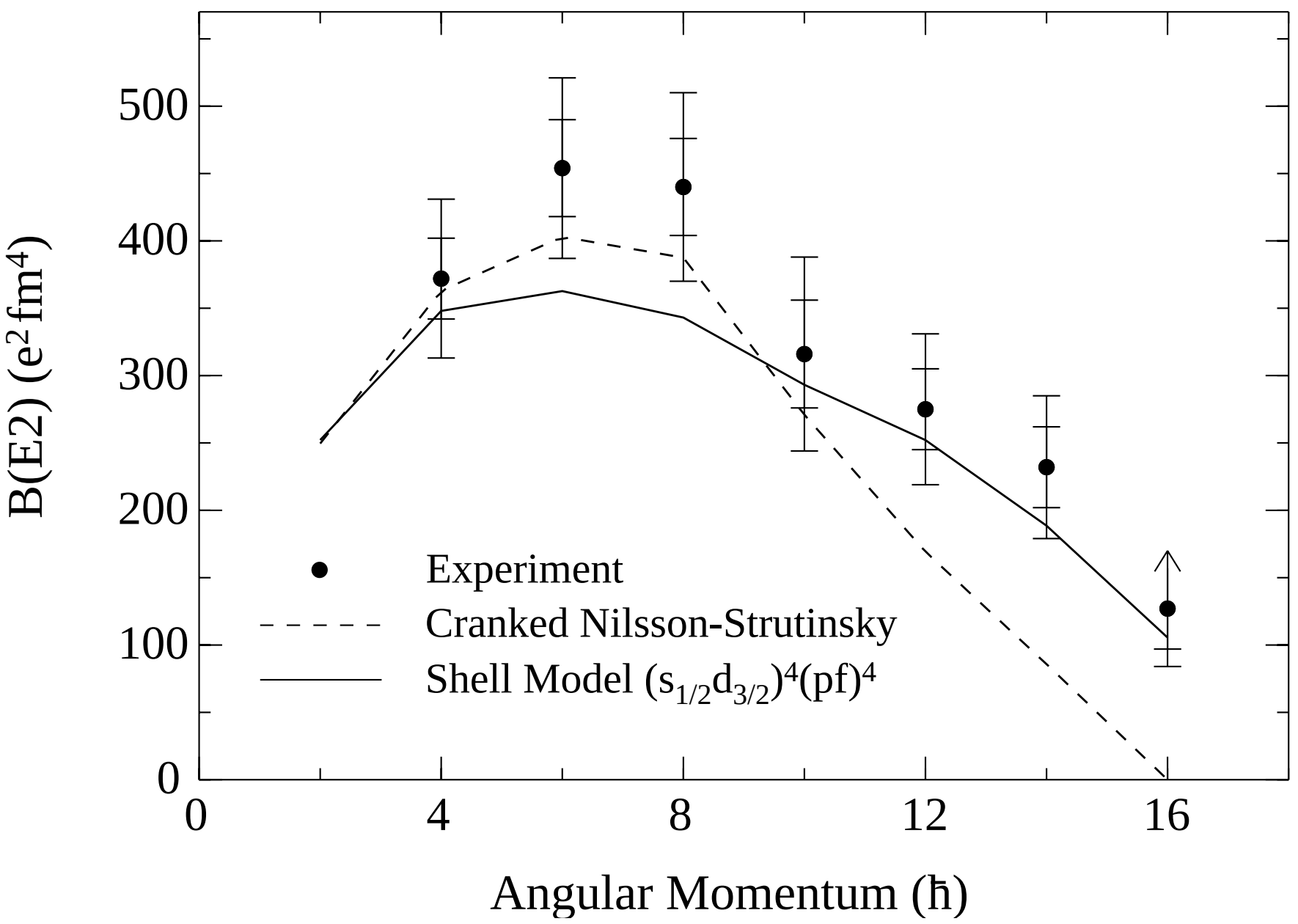

Fig. 6. Measured $B(E2; I \to I-2)$ values for the SD band in $^{36}$Ar compared with the results of cranked Nilsson-Strutinsky (dashed line) and $(s_{1/2}d_{3/2})^4(pf)^4$ spherical shell model (solid line) calculations. The inner error bars represent the statistical uncertainties, while the outer error bars show the effects of systematic $\pm 10\%$ shifts in the stopping powers.

Figure 7 shows the occupancies of spherical $j$-shells in the SD band from the CNS and SM calculations. We note that the relatively constant occupancies at low spin confirm the concept of a rotational band built on a fixed intrinsic configuration. At higher spins, the increasingly rapid occupancy changes reflect the changes in the intrinsic deformation shown in Fig. 3. For the $pf$ shell, the occupancies predicted by the two calculations are very similar, indicating the dominance of the quadrupole deformation in determining the structure of the wavefunction for this highly collective rotational band. The full treatment of all correlations, in particular pairing, in the shell model leads only to a small increase in the occupancies of the upper $pf$ shell orbitals relative to the dominant deformation-induced mixing of the spherical $j$-shells reflected in the CNS occupancies. For the

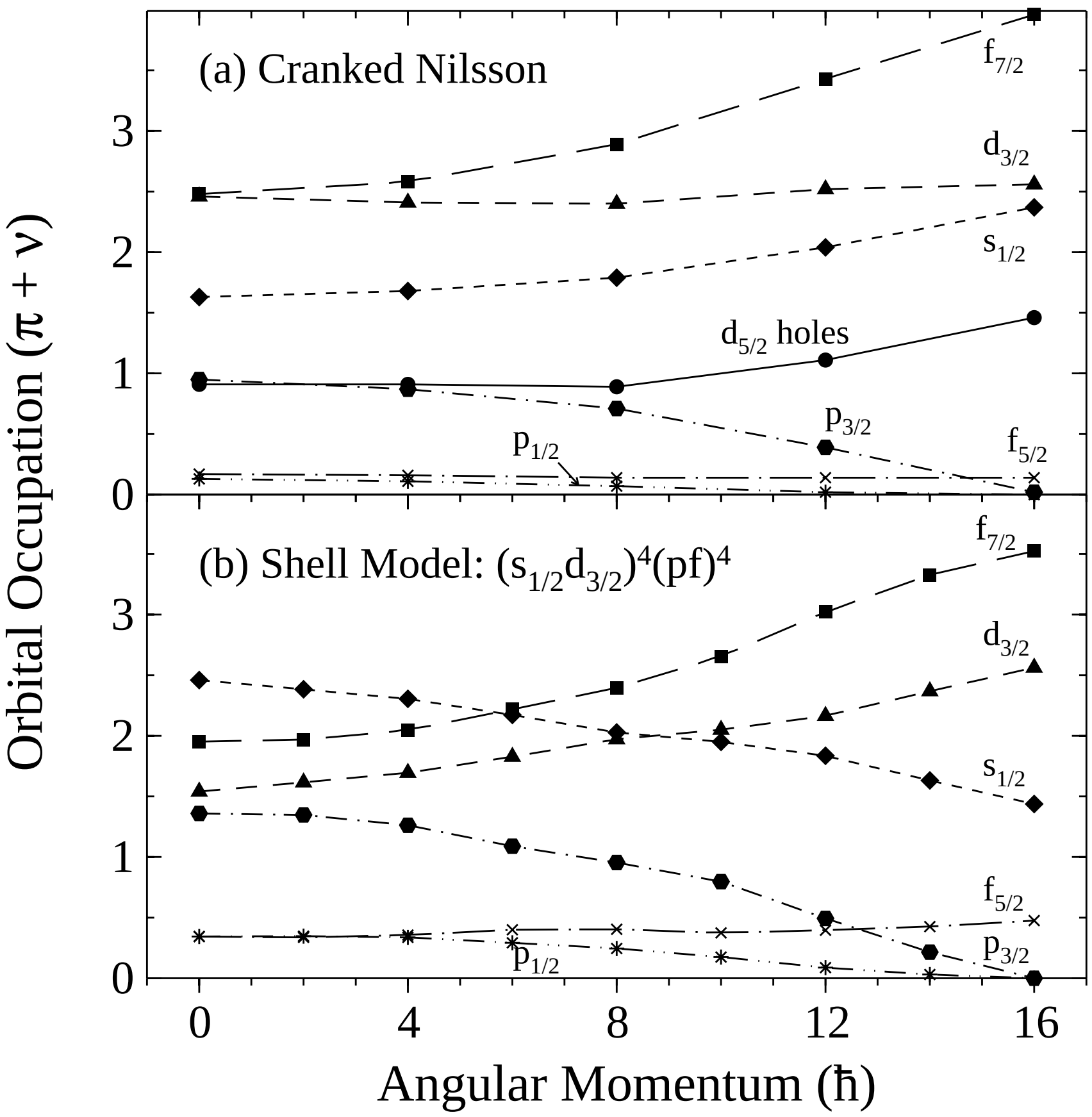

Fig. 7. Occupancies of spherical $j$-shells in the $^{36}$Ar superdeformed band from (a) cranked Nilsson-Strutinsky and (b) $(s_{1/2}d_{3/2})^4(pf)^4$ shell model calculations.

*sd* shell, dimensionality considerations required a truncation of the shell model space to $s_{1/2}d_{3/2}$. The CNS calculations, however, indicate that the $d_{5/2}$ orbital is only $\sim$ 90% full in the SD band. It is worth noting that the intrinsic quadrupole moment $Q_0 = 113$ $e$ fm$^2$ deduced for the SD band from the $(s_{1/2}d_{3/2})^4(pf)^4$ SM calculation is already 80% of the maximum possible in the full $(sd)^{16}(pf)^4$ space (as attained in the unrealistic SU(3) limit). Nevertheless, a $\sim$ 5–10% increase in $Q_0$ due to the $d_{5/2}$-hole component of the wavefunction is plausible, and would account for the $\sim$ 10–20% underestimate of the $B(E2)$ values by the current shell model calculations. Further study of the role of the $d_{5/2}$ orbital in the structure of this highly collective band, perhaps through Monte Carlo Shell Model techniques [16], would clearly be desirable.

## 4. Conclusions

A superdeformed rotational band has been identified in the light $N = Z$ nucleus $^{36}$Ar. The linking of this band to previously known low-spin states, combined with lifetime measurements throughout the band, have established absolute excitation energies, spins, and parities, as well as in-band and out-of-band $B(E2)$ values for the SD states. With effectively complete spectroscopic information and a model space large enough for substantial collectivity to develop, yet small enough to be approached from the shell-model perspective, the $^{36}$Ar SD band offers many exciting opportunities for further studies of the microscopic structure of collective rotational motion in nuclei.

## 5. Acknowledgements

This work has been partially supported by the Natural Sciences and Engineering Research Council of Canada, the U.S. D.O.E. under Contracts DE-AC03-76SF00098, W-31-109-ENG-38, and DE-FG05-88ER40406, the Swedish Institute, NFR (Sweden), DGES (Spain) under Grant No. PB96-53, and IN2P3 (France).